\documentclass[12pt,fleqn]{article}
\usepackage{amsfonts}
\usepackage{amssymb}
\usepackage{amsmath}
\usepackage{geometry}
\usepackage{setspace}
\usepackage{caption}

\input{tcilatex}
\begin{document}

\title{Artificial Intelligence as Structural Estimation:\\
Economic Interpretations of Deep Blue, Bonanza, and AlphaGo\thanks{%
First version: October 30, 2017. This paper benefited from seminar comments
at Riken AIP, Georgetown, Tokyo, Osaka, Harvard, Johns Hopkins, and \textit{%
The Third Cambridge Area Economics and Computation Day} conference at
Microsoft Research New England, as well as conversations with Susan Athey,
Xiaohong Chen, Jerry Hausman, Greg Lewis, Robert Miller, Yusuke Narita, Aviv
Nevo, Anton Popov, John Rust, Takuo Sugaya, Elie Tamer, and Yosuke Yasuda.}}
\author{Mitsuru Igami\thanks{%
Yale Department of Economics and MIT Department of Economics. E-mail:
mitsuru.igami@gmail.com.}}
\date{March 1, 2018}
\maketitle

\begin{abstract}
Artificial intelligence (AI) has achieved superhuman performance in a
growing number of tasks, but understanding and explaining AI remain
challenging. This paper clarifies the connections between machine-learning
algorithms to develop AIs and the econometrics of dynamic structural models
through the case studies of three famous game AIs. Chess-playing Deep Blue
is a calibrated value function, whereas shogi-playing Bonanza is an
estimated value function via Rust's (1987) nested fixed-point method.
AlphaGo's \textquotedblleft supervised-learning policy
network\textquotedblright \ is a deep neural network implementation of Hotz
and Miller's (1993) conditional choice probability estimation; its
\textquotedblleft reinforcement-learning value network\textquotedblright \
is equivalent to Hotz, Miller, Sanders, and Smith's (1994) conditional
choice simulation method. Relaxing these AIs' implicit econometric
assumptions would improve their structural interpretability.

\bigskip


\bigskip

\end{abstract}

\clearpage

\section{Introduction}

Artificial intelligence (AI) has achieved human-like performance in a
growing number of tasks, such as visual recognition and natural language
processing.\footnote{%
The formal definition of AI seems contentious, partly because scholars have
not agreed on the definition of intelligence in the first place. This paper
follows a broad definition of AI as computer systems able to perform tasks
that traditionally required human intelligence.} The classical games of
chess, shogi (Japanese chess), and Go were once thought to be too
complicated and intractable for AI, but computer scientists have overcome
these challenges. In chess, IBM's computer system named Deep Blue defeated
Grandmaster Garry Kasparov in 1997. In shogi, a machine-learning-based
program called Bonanza challenged (and was defeated by) Ry\={u}\={o}
champion Akira Watanabe in 2007, but one of its successors (Ponanza) played
against Meijin champion Amahiko Satoh and won in 2017. In Go, Google
DeepMind developed AlphaGo, a deep-learning-based program, which beat the
2-dan European champion Fan Hui in 2015, a 9-dan (highest rank) professional
Lee Sedol in 2016, and the world's best player Ke Jie in 2017.

Despite such remarkable achievements, one of the lingering criticisms of AI
is its lack of transparency. The internal mechanism seems like a black box
to most people, including the human experts of the relevant tasks,\footnote{%
For example, Yoshiharu Habu, the strongest shogi player in recent history,
states he does not understand certain board-evaluation functions of computer
shogi programs (Habu and NHK [2017]).} which raises concerns about
accountability and responsibility. The desire to understand and explain the
functioning of AI is not limited to the scientific community. For example,
the US Department of Defense airs its concern that \textquotedblleft the
effectiveness of these systems is limited by the machine's current inability
to explain their decisions and actions to human users,\textquotedblright \
which led it to host the Explainable AI (XAI) program aimed at developing
\textquotedblleft understandable\textquotedblright \ and \textquotedblleft
trustworthy\textquotedblright \ machine learning.\footnote{%
See https://www.darpa.mil/program/explainable-artificial-intelligence
(accessed on October 17, 2017).}

This paper examines three prominent game AIs in recent history: Deep Blue,
Bonanza, and AlphaGo. I have chosen to study this category of AIs because
board games represent an archetypical task that has required human
intelligence, including cognitive skills, decision-making, and
problem-solving. They are also well-defined problems for which economic
interpretations are more natural than for, say, visual recognition and
natural language processing. The main finding from this paper's case studies
is that these AIs' key components are mathematically equivalent to
well-known econometric methods to estimate dynamic structural models.

Chess experts and IBM's engineers manually adjusted thousands of parameters
in Deep Blue's \textquotedblleft evaluation function,\textquotedblright \
which quantifies the probability of eventual winning as a function of the
current positions of pieces (i.e., state of the game) and therefore could be
interpreted as an approximate value function. Deep Blue is a calibrated
value function with a linear functional form.

By contrast, the developer of Bonanza constructed a dataset of professional
shogi games, and used a discrete-choice regression and a backward-induction
algorithm to determine the parameters of its value function. Hence, his
method of \textquotedblleft supervised learning\textquotedblright \ is
equivalent to Rust's (1987) nested fixed-point (NFXP) algorithm, which
combined a discrete-choice model with dynamic programming (DP) in the
maximum likelihood estimation (MLE) framework. Bonanza is an empirical model
of human shogi players that is estimated by this direct (or
\textquotedblleft full-solution\textquotedblright ) method.

Google DeepMind's AlphaGo (its original version) embodies an alternative
approach to estimating dynamic structural models: two-step estimation.%
\footnote{%
This paper focuses on the original version of AlphaGo,\ published in 2016,
and distinguishes it from its later version, \textquotedblleft AlphaGo
Zero,\textquotedblright \ published in 2017. The latter version contains few
econometric elements, and is not an immediate subject of my case study,
although I discuss some of its interesting features in section 5.} Its first
component, the \textquotedblleft supervised-learning (SL) policy
network,\textquotedblright \ predicts the moves of human experts as a
function of the board state. It is an empirical policy function with a class
of nonparametric basis functions (DNN: deep neural network) that is
estimated by MLE, using data from online Go games. Thus, the SL policy
network is a DNN implementation of Hotz and Miller's (1993) first-stage
conditional choice probability (CCP) estimation.

AlphaGo's value function, called \textquotedblleft reinforcement-learning
(RL) value network,\textquotedblright \ is constructed by simulating many
games based on the self-play of the SL policy network and estimating another
DNN model that maps state to the probability of winning. This procedure is
equivalent to the second-stage conditional choice simulation (CCS)
estimation, proposed by Hotz, Miller, Sanders, and Smith (1994) for
single-agent DP, and by Bajari, Benkard, and Levin (2007) for dynamic games.

Thus, these leading game AIs and the core algorithms for their development
turn out to be successful applications of the empirical methods to implement
dynamic structural models. After introducing basic notations in section 2, I
describe the main components of Deep Blue, Bonanza, and AlphaGo in sections
3, 4, and 5, respectively, and explain their structural interpretations.
Section 6 clarifies some of the implicit assumptions underlying these AIs,
such as (the absence of) unobserved heterogeneity, strategic interactions,
and various constraints human players are facing in real games. Section 7
concludes by suggesting that relaxing some of these assumptions and
explicitly incorporating more realistic features of the data-generating
process could help make AIs both more human-like (if needed) and more
amenable to structural interpretations.

\textbf{Literature }This paper clarifies the equivalence between some of the
algorithms for developing game AI and the aforementioned econometric methods
for estimating dynamic models. As such, the most closely related papers are
Rust (1987), Hotz and Miller (1993), and Hotz, Miller, Sanders, and Smith
(1994). The game AIs I analyze in this paper are probably the most
successful (or at least the most popular) empirical applications of these
methods. For a historical review of numerical methods for dynamic
programming, see Rust (2017).

At a higher level, the purpose of this paper is to clarify the connections
between machine learning and econometrics in certain areas. Hence, the paper
shares the spirit of, for example, Belloni, Chernozhukov, and Hansen (2014),
Varian (2014), Athey (2017), and Mullainathan and Spiess (2017), among many
others in the rapidly growing literature on data analysis at the
intersection of computer science and economics.

\section{Model}

\subsubsection*{Rules}

Chess, shogi, and Go belong to the same class of games, with two players ($%
i=1,2$), discrete time ($t=1,2,...$), alternating moves (players 1 and 2
choose their actions, $a_{t}$, in odd and even periods, respectively),
perfect information, and deterministic state transition, 
\begin{equation}
s_{t+1}=f\left( s_{t},a_{t}\right) ,  \label{eq - transition}
\end{equation}%
where both the transition, $f\left( \cdot \right) $, and the initial state, $%
s_{1}$, are completely determined by the rule of each game.\footnote{%
This setup abstracts from the time constraints in official games because the
developers of game AIs typically do not incorporate them at the
data-analysis stage. Hence, $t$ represents turn-to-move, not clock time.
Section 6 investigates this issue.}

Action space is finite and is defined by the rule as \textquotedblleft legal
moves,\textquotedblright 
\begin{equation}
a_{t}\in \mathcal{A}\left( s_{t}\right) .  \label{eq- action space}
\end{equation}%
State space is finite as well, and consists of four mutually exclusive
subsets:%
\begin{equation}
s_{t}\in \mathcal{S}=\mathcal{S}_{cont}\sqcup \mathcal{S}_{win}\sqcup 
\mathcal{S}_{loss}\sqcup \mathcal{S}_{draw},  \label{eq - state space}
\end{equation}%
where I denote \textquotedblleft win\textquotedblright \ and
\textquotedblleft loss\textquotedblright \ from the perspective of player 1
(e.g., player 1 wins and player 2 loses if $s_{t}\in \mathcal{S}_{win}$).
Neither player wins if $s_{t}\in \mathcal{S}_{draw}$. The game continues as
long as $s_{t}\in \mathcal{S}_{cont}$.

The two players' payoffs sum to zero:%
\begin{equation}
u_{1}\left( s_{t}\right) =\left \{ 
\begin{array}{ll}
1 & \text{if }s_{t}\in \mathcal{S}_{win}, \\ 
-1 & \text{if }s_{t}\in \mathcal{S}_{loss},\text{ and} \\ 
0 & \text{otherwise,}%
\end{array}%
\right.  \label{eq - utility}
\end{equation}%
with $u_{2}\left( s_{t}\right) $ defined in a similar manner (but with
win/loss payoffs flipped). This setup means chess, shogi, and Go are
well-defined finite games. In principle, such games can be solved exactly
and completely by backward induction from the terminal states.

In practice, even today's supercomputers and a cloud of servers cannot solve
them within our lifetime, because the size of the state space, $\left \vert 
\mathcal{S}\right \vert $, is large. The approximate $\left \vert \mathcal{S}%
\right \vert $ of chess, shogi, and Go are $10^{47}$, $10^{71}$, and $10^{171}
$, respectively, which are comparable to the number of atoms in the
observable universe ($10^{78}\sim 10^{82}$) and certainly larger than the
total information-storage capacity of humanity (in the order of $10^{20}$
bytes).\footnote{%
In 2016, the world's hard disk drive (HDD) industry produced a total of 693
exabytes (EB), or $6.93\times 10^{20}$ bytes.}

\section{Chess: Deep Blue}

\subsection{Algorithms}

IBM's Deep Blue is a computer system with custom-built hardware and software
components. I focus on the latter, programming-related part. Deep Blue's
program consists of three key elements: an evaluation function, a search
algorithm, and databases.

\subsubsection*{Evaluation Function}

The \textquotedblleft evaluation function\textquotedblright \ of Deep Blue is
a linear combination of certain features of the current board state $s_{t}$.
It quantifies the probability of eventual winning ($\Pr_{win}$) or its
monotonic transformation, $g\left( \Pr_{win}\right) $:%
\begin{equation}
V_{DB}\left( s_{t};\theta \right) =\theta _{1}x_{1,t}+\theta
_{2}x_{2,t}+\cdots +\theta _{K}x_{K,t},  \label{eq - V_DB}
\end{equation}%
where $\theta \equiv \left( \theta _{1},\theta _{2},\ldots ,\theta
_{K}\right) $ is a vector of $K$ parameters and $x_{t}\equiv \left(
x_{1,t},x_{2,t},\ldots ,x_{K,t}\right) $ is a vector of $K$ observable
characteristics of $s_{t}$. The published version featured $K=8,150$
parameters (Campbell, Hoane, and Hsu [2002]).

A typical evaluation function for computer chess considers the
\textquotedblleft material value\textquotedblright \ associated with each
type of piece, such as 1 point for a pawn, 3 points for a knight, 3 points
for a bishop, 5 points for a rook, 9 points for a queen, and an arbitrarily
many points for a king (e.g., 200 or 1 billion), because the game ends when
a king is captured. Other factors include the relative positions of these
pieces, such as pawn structure, protection of kings, and experts' opinion
that a pair of bishops are usually worth more than the sum of their
individual material values. Finally, the importance of these factors may
change depending on the phase of the game: opening, middle, or endgame.

Reasonable parameterization and the choice of board characteristics
(variables) require expert knowledge. Multiple Grandmasters (the highest
rank of chess players) advised the Deep Blue development team. However, they
did not use statistical analysis or data from professional players' games.
In other words, they did not estimate $\theta $. Each of the 8,150
parameters, $\theta $, was \textit{manually adjusted} until the program's
performance reached a satisfactory level.

\subsubsection*{Search Algorithm}

The second component of Deep Blue is \textquotedblleft
search,\textquotedblright \ or a solution algorithm to choose the optimal
action at each turn to move. In its \textquotedblleft full-width
search\textquotedblright \ procedure, the program evaluates every possible
position for a fixed number of future moves along the game tree, using the
\textquotedblleft minimax algorithm\textquotedblright \ and some
\textquotedblleft pruning\textquotedblright \ methods. This
\textquotedblleft search\textquotedblright \ is a version of numerical
backward induction.

\subsubsection*{Databases}

Deep Blue uses two databases: one for the endgame and the other for the
opening phase.

The endgame database embodies the cumulative efforts by the computer chess
community to solve the game in an exact manner. Ken Thompson and Lewis
Stiller developed Deep Blue's endgame database with all five-piece positions
(i.e., the states with only five pieces, and all possible future states that
can be reached from them), as well as selected six-piece positions.\footnote{%
A database with all five-piece endings requires 7.05 gigabytes of hard disk
space; storing all six-piece endings requires 1.2 terabytes.}

The second database is an \textquotedblleft opening book,\textquotedblright \
which is a collection of common openings (i.e., move patterns at the
beginning of the game) experts consider good plays. It also contains good
ways to counter the opponent's poor openings, again based on the judgment by
experts. Grandmasters Joel Benjamin, Nick De Firmian, John Fedorowicz, and
Miguel Illescas created one with about 4,000 positions, by hand.

The team also used some data analysis to prepare an \textquotedblleft
extended book\textquotedblright \ to guard against non-standard opening
positions. Campbell, Hoane, and Hsu's (2002) description suggests they
constructed a parametric move-selection function based on the data analysis
of 700,000 human games. They even incorporated annotators' commentary on the
moves. Nevertheless, the use of data analysis seems limited to this back-up
database.

\subsubsection*{Performance}

Deep Blue lost a match to the top-ranked Grandmaster Garry Kasparov in 1996,
but defeated him in 1997. Since then, the use of computer programs has
become widespread in terms of both training and games (e.g., those played by
hybrid teams of humans and computers).\footnote{%
See Kasparov (2007), for example.}

\subsection{Deep Blue Is a Calibrated Value Function}

The fact that IBM manually adjusted the parameter $\theta $ by trial and
error means Deep Blue is a fruit of painstaking efforts to \textquotedblleft
calibrate\textquotedblright \ a value function with thousands of parameters:
Deep Blue is a calibrated value function.

A truly optimal value function would obviate the need for any
forward-looking and backward-induction procedures to solve the game (i.e.,
the \textquotedblleft search algorithm\textquotedblright ), because the true
value function should embody such solution. However, the parametric value
function is merely an attempt to approximate the optimal one. Approximation
errors (or misspecification) seem to leave room for performance improvement
by the additional use of backward induction, although such benefits are not
theoretically obvious.

The \textquotedblleft full-width search\textquotedblright \ procedure is a
brute-force numerical search for the optimal choice by backward induction,
but solving the entire game is infeasible. Hence, this backward induction is
performed on a \textit{truncated} version of the game tree (truncated at
some length $L$ from the current turn $t$). The \textquotedblleft
terminal\textquotedblright \ values at turn $\left( t+L\right) $ are given by
the parametric value function (\ref{eq - V_DB}). The opponent is assumed to
share the same terminal-value function at $\left( t+L\right) $ and to choose
its move to minimize the focal player's $V_{DB}\left( s_{t+L};\theta \right) 
$.

Thus, Deep Blue\ is a parametric (linear) function to approximate the
winning probability at the end of a truncated game tree, $V_{DB}\left(
s_{t+L};\theta \right) $, in which the opponent shares exactly the same
value function and the time horizon. In other words, Deep Blue is a
calibrated, approximate terminal-value function in a game the program plays
against its doppelg\"{a}nger.

\section{Shogi: Bonanza}

\subsection{Algorithm}

In 2005, Kunihito Hoki, a chemist who specialized in computer simulations at
the University of Toronto, spent his spare time developing a shogi-playing
program named Bonanza, which won the world championship in computer shogi in
2006. Hoki's Bonanza revolutionized the field of computer shogi by
introducing machine learning to \textquotedblleft train\textquotedblright \
(i.e., estimate) a more flexible evaluation function than either those for
chess or those designed for the existing shogi programs.\footnote{%
Hoki acknowledges machine-learning methods had previously been used in
computer programs to play Backgammon and Reversi (\textquotedblleft
Othello\textquotedblright ), but says he could not find any successful
applications to chess or shogi in the literature (Hoki and Watanabe [2007]).}

\subsubsection*{More Complicated Evaluation Function}

The developers at IBM could manually adjust 8,150 parameters in $%
V_{DB}\left( s_{t};\theta \right) $\ and beat the human chess champions. The
same approach did not work for shogi. Shogi programs before Bonanza could
only compete at amateurs' level at best. This performance gap between chess
and shogi AIs is rooted in the more complicated state space of shogi, with $%
\left \vert \mathcal{S}_{shogi}\right \vert \approx 10^{71}>10^{47}\approx
\left \vert \mathcal{S}_{chess}\right \vert $.

Several factors contribute to the complexity of shogi: a larger board size ($%
9\times 9>8\times 8$), more pieces ($40>32$), and more types of pieces ($8>6$%
). In addition, most of the pieces have limited mobility,\footnote{%
Four of the eight types of pieces in shogi can move only one unit distance
at a time, whereas only two of the six types of pieces in chess (pawn and
king) have such low mobility. The exact positions of pieces becomes more
important in characterizing the state space when mobility is low, whereas
high mobility makes pure \textquotedblleft material
values\textquotedblright \ relatively more important, because pieces can be
moved to wherever they are needed within a few turns.} and, other than
kings, never die. Non-king pieces are simply \textquotedblleft
captured,\textquotedblright \ not killed, and can then be \textquotedblleft
dropped\textquotedblright \ (re-deployed on the capturer's side) almost
anywhere on the board at any of the capturer's subsequent turns. This last
feature is particularly troublesome for an attempt to solve the game
exactly, because the effective $\left \vert \mathcal{S}\right \vert $\ does
not decrease over time.

Hoki designed a flexible evaluation function by factorizing the positions of
pieces into (i) the positions of any three pieces including the kings and
(ii) the positions of any three pieces including only one king. This
granular characterization turned out to capture important features of the
board through the relative positions of three-piece combinations. Bonanza's
evaluation function, $V_{BO}\left( s_{t};\theta \right) $, also incorporated
other, more conventional characteristics, such as individual pieces'
material values (see Hoki and Watanabe [2007], pp. 119--120, for details).
As a result, $V_{BO}\left( s_{t};\theta \right) $ has a linear functional
form similar to (\ref{eq - V_DB}) but contains $K=50$ $million$ variables
and the same number of parameters (Hoki [2012]).

\subsubsection*{Machine Learning (Logit-like Regression)}

That the Deep Blue team managed to adjust thousands of parameters for the
chess program by human hands is almost incredible. But the task becomes
impossible with 50 million parameters. Hoki gave up on manually tuning
Bonanza's $\theta $ and chose to use statistical methods to automatically
adjust $\theta $ based on the data from the professional shogi players'
50,000 games on official record: supervised learning.

Each game takes 100 moves on average. Hence, the data contain approximately
5 million pairs of $\left( a_{t},s_{t}\right) $.\footnote{%
In earlier versions of Bonanza, Hoki also used additional data from 50,000
unofficial, online game records as well, to cover some rare states such as
nyuu-gyoku positions (in which a king enters the opponent's territory and
becomes difficult to capture, because the majority of shogi pieces can only
move forward, not backward). However, he found the use of data from amateur
players' online games weakened Bonanza's play, and subsequently abandoned
this approach (Hoki [2012]).} The reader might notice the sample size is
smaller than $\left \vert \theta \right \vert $ (50 million). Hoki reduced the
effective number of parameters by additional restrictions to
\textquotedblleft stabilize the numerical optimization
process.\textquotedblright 

Like Deep Blue, Bonanza chooses its move at each of its turns $t$ by
searching for the action $a_{t}$ that maximizes $V_{BO}$ in some future turn 
$t+L$,%
\begin{equation}
a_{t}^{\ast }=\arg \max_{a\in \mathcal{A}\left( s_{t}\right) }\left \{
V_{BO}\left( s_{t+L};\theta \right) \right \} ,  \label{eq - Bonanza action}
\end{equation}%
assuming the opponent shares the same terminal-value function and tries to
minimize it. Note the \textquotedblleft optimal\textquotedblright \ choice $%
a_{t}^{\ast }$ is inherently related to $\theta $ through the objective
function $V_{BO}\left( s_{t+L};\theta \right) $. This relationship can be
exploited to infer $\theta $ from the data on $\left( a_{t},s_{t}\right) $.
Hoki used some variant of the discrete-choice regression method to determine
the values of $\theta $.\footnote{%
Tomoyuki Kaneko states he also used some machine-learning methods as early
as 2003 for his program named GPS Shogi (Kaneko [2012]). Likewise, Yoshimasa
Tsuruoka writes he started using logit regressions in 2004 for developing
his own program, Gekisashi (Tsuruoka [2012]). But shogi programmers seem to
agree that Hoki's Bonanza was the first to introduce data-analysis methods
for constructing an evaluation function in a wholesale manner.}

\subsubsection*{Performance}

Bonanza won the world championship in computer shogi in 2006 and 2013. In
2007, the Ry\={u}\={o} (\textquotedblleft dragon king,\textquotedblright \
one of the two most prestigious titles) champion Akira Watanabe agreed to
play against Bonanza and won. After the game, however, he said he regretted
agreeing to play against it, because he felt he could have lost with
non-negligible probabilities. Hoki made the source code publicly available.
The use of data and machine learning for computer shogi was dubbed the
\textquotedblleft Bonanza method\textquotedblright \ and copied by most of
the subsequent generations of shogi programs.

Issei Yamamoto, a programmer, named his software Ponanza out of respect for
the predecessor, claiming his was a lesser copy of Bonanza. From 2014,
Ponanza started playing against itself in an attempt to find
\textquotedblleft stronger\textquotedblright \ parameter configurations than
those obtained (estimated) from the professional players' data:
reinforcement learning (Yamamoto [2017]). Eventually, Ponanza became the
first shogi AI to beat the Meijin (\textquotedblleft
master,\textquotedblright \ the other most prestigious title) champion in
2017, when Amahiko Satoh lost two straight games.

\subsection{Structural Interpretation: Bonanza Is Harold Zurcher}

Bonanza is similar to Deep Blue. Its main component is an approximate
terminal-value function, and the \textquotedblleft optimal\textquotedblright
\ action is determined by backward induction on a truncated game tree of
self play (equation \ref{eq - Bonanza action}). The only difference is the
larger number of parameters (50 million), which reflects the complexity of
shogi and precludes any hopes for calibration. Hence, Hoki approached the
search for $\theta $ as a data-analysis problem.

Accordingly, Bonanza is an empirical model of professional shogi players, in
the same sense that Rust (1987) is an empirical model of Harold Zurcher, a
Madison, Wisconsin, city-bus superintendent. Rust studied his record of
engine-replacement decisions and estimated his utility function, based on
the principle of revealed preference. This comparison is not just a
metaphor. The machine-learning algorithm to develop Bonanza is almost
exactly the same as the structural econometric method to estimate an
empirical model of Harold Zurcher.

Rust's (1987) \textquotedblleft full-solution\textquotedblright \ estimation
method consists of two numerical optimization procedures that are nested.
First, the overall problem is to find $\theta $ that makes the model's
prediction $a_{t}^{\ast }$ (as a function of $s_{t}$ and $\theta $) fit the
observed action-state pairs in the data $\left( a_{t},s_{t}\right) $.
Second, the nested sub-routine takes particular $\theta $ as an input and
solves the model to find the corresponding policy function (optimal
strategy),%
\begin{equation}
a_{t}^{\ast }=\sigma \left( s_{t};V_{BO}\left( \cdot ;\theta \right) \right)
.  \label{eq - optimal strategy}
\end{equation}%
The first part is implemented by the maximum likelihood method (i.e., the
fit is evaluated by the proximity between the observed and predicted choice
probabilities). The second part is implemented by value-function iteration,
that is, by numerically solving a contraction-mapping problem to find a
fixed point, which is guaranteed to exist and is unique for a well-behaved
single-agent dynamic programming (DP) problem. This algorithm is called
nested fixed-point (NFXP) because of this design.

Hoki developed Bonanza in the same manner. The overall problem is to find $%
\theta $ that makes Bonanza predict the human experts' actions in the data (%
\ref{eq - optimal strategy}). The nested sub-routine takes $\theta $ as an
input and numerically searches for the optimal action $a_{t}^{\ast }$ by
means of backward induction. The first part is implemented by logit-style
regressions (i.e., the maximum-likelihood estimation of the discrete-choice
model in which the error term is assumed i.i.d. type-1 extreme value). This
specification is the same as Rust's. The second part proceeds on a truncated
game tree, whose \textquotedblleft leaves\textquotedblright \ (i.e., terminal
values at $t+L$) are given by the approximate value function $V_{BO}\left(
s_{t+L};\theta \right) $, and the opponent is assumed to play the same
strategy as itself:%
\begin{equation}
\sigma _{-i}=\sigma _{i}.  \label{eq - same strategy}
\end{equation}

Strictly speaking, Bonanza differs from (the empirical model of) Harold
Zurcher in two aspects. Bonanza plays a two-player game with a finite
horizon, whereas Harold Zurcher solves a single-agent DP with an infinite
horizon. Nevertheless, these differences are not as fundamental as one might
imagine at first glance, because each of them can be solved for a unique
\textquotedblleft optimal\textquotedblright \ strategy $\sigma ^{\ast }$ in
the current context. An alternating-move game with a finite horizon has a
unique equilibrium when i.i.d. utility shock is introduced and breaks the
tie between multiple discrete alternatives. Igami (2017, 2018) demonstrates
how Rust's NFXP naturally extends to such cases with a deterministic order
of moves; Igami and Uetake (2017) do the same with a stochastic order of
moves. Thus, Bonanza is to Akira Watanabe what Rust (1987) is to Harold
Zurcher.

\section{Go: AlphaGo}

\subsection{Algorithm}

The developers of AIs for chess and shogi had successfully parameterized
state spaces and constructed evaluation functions. Meanwhile, the developers
of computer Go struggled to find any reasonable parametric representation of
the board.

Go is more complicated than chess and shogi, with $\left \vert \mathcal{S}%
_{go}\right \vert \approx 10^{171}>10^{71}\approx \left \vert \mathcal{S}%
_{shogi}\right \vert $. Go has only one type of piece, a stone, and the goal
is to occupy larger territories than the opponent when the board is full of
black and white stones (for players 1 and 2, respectively). However, the $%
19\times 19$ board size is much larger, and so is the action space.
Practically all open spaces constitute legal moves. The local positions of
stones seamlessly interact with the global ones. Even the professional
players cannot articulate what distinguishes good positions from bad ones,
frequently using phrases that are ambiguous and difficult to codify. The
construction of a useful evaluation function was deemed impossible.

Instead, most of the advance since 2006 had been focused on improving
game-tree search algorithms (Yoshizoe and Yamashita [2012], Otsuki [2017]).
Even though the board states in the middle of the game are difficult to
codify, the terminal states are unambiguous, with either win or loss.
Moreover, a \textquotedblleft move\textquotedblright \ in Go does not involve
moving pieces that are already present on the board; it comprises simply
dropping a stone on an open space from outside the board. These features of
Go make randomized \textquotedblleft play-out\textquotedblright \ easy. That
is, the programmer can run Monte Carlo simulations in which black and white
stones are alternatingly dropped in random places until the board is filled.
Repeat this forward simulation many times, and one can calculate the
probability of winning from any arbitrary state of the board $s_{t}$. This
basic idea is behind a method called Monte Carlo tree search (MCTS).

Given the large state space, calculating the probability of winning (or its
monotonic transformation $V\left( s_{t}\right) $) for all $s_{t}$'s remains
impractical. However, a computer program can use this approach \textit{in
real time} to choose the next move, because it needs to compare only $%
\left \vert \mathcal{A}\left( s_{t}\right) \right \vert <361=19\times 19$
alternative actions and their immediate consequences (i.e., states) at its
turn to move. Forward simulations involve many calculations, but each
operation is simple and parallelizable. That is, computing one history of
future play does not rely on computing another history. Likewise,
simulations that start from a particular state $s_{t+1}=f\left(
s_{t},a\right) $ do not have to wait for the completion of other simulations
that start from $s_{t+1}^{\prime }=f\left( s_{t},a^{\prime }\right) $, where 
$a\neq a^{\prime }$. Such computational tasks can be performed
simultaneously on multiple computers, processors, cores, or GPUs (graphic
processing units). If the developer can use many computers during the game,
the MCTS-based program can perform sufficiently many numerical operations to
find good moves within a short time.

Thus, MCTS was the state of the art in computer Go programming when Demis
Hassabis and his team at Google DeepMind proposed a deep-learning-based AI,
AlphaGo. The four components of AlphaGo are (i) a policy network, (ii) RL,
(iii) a value network, and (iv) MCTS. Among these four components, (i) and
(iii) were the most novel features relative to the existing game AIs.

\subsubsection*{Component 1: Supervised Learning (SL) of the Policy Network}

The first component of AlphaGo is a \textquotedblleft policy
network,\textquotedblright \ which is a deep neural network (DNN) model to
predict professional players' move $a_{t}$ as a function of current state $%
s_{t}$. It is a policy function as in (\ref{eq - optimal strategy}), with a
particular functional form and 4.6 million \textquotedblleft
weights\textquotedblright \ (i.e., parameter vector $\psi $).\footnote{%
I use $\psi $ to denote the parameter vector of policy function, and
distinguish it from the parameter vector of the value function $\theta $.}

Like Hoki did for Bonanza, the AlphaGo team determined $\psi $ by using the
data from an online Go website named Kiseido Go Server (KGS). Specifically,
they used the KGS record on 160,000 games played by high-level (6-9 dan)
professionals. A game lasts 200 moves on average, and eight symmetric
transformations (i.e., rotations and flipping) of the board generate
nominally different states. Hence, the effective size of the dataset is%
\begin{eqnarray*}
256\text{ million (action-state pairs)} &=&160,000\text{ (games)}\times 200%
\text{ (moves/game)} \\
&&\times 8\text{ (symmetric transformations).}
\end{eqnarray*}%
Note the sample size is still small (negligible) relative to $\left \vert 
\mathcal{S}_{go}\right \vert \approx 10^{171}$.

Its basic architecture is a standard Convolutional Neural Network (CNN),
which is known to perform well in image-recognition tasks, among others. The
Appendix describes the details of AlphaGo's DNN functional-form
specification. Its final output is the prediction of choice probabilities
across all legal moves based on the following logit formula:%
\begin{equation}
\Pr \left( a_{t}=j|s_{t};\psi \right) =\frac{\exp \left( y_{j}\left(
s_{t};\psi \right) \right) }{\sum_{j^{\prime }\in \mathcal{A}\left(
s_{t}\right) }\exp \left( y_{j^{\prime }}\left( s_{t};\psi \right) \right) },
\label{eq - logit}
\end{equation}%
where $j$ indexes actions and $y_{j}\left( s_{t};\psi \right) $ is the
deterministic part of the latent value of choosing action $j$ in state $s_{t}
$ given parameter $\psi $.

This specification is \textquotedblleft deep\textquotedblright \ in the sense
that the model contains multiple layers, through which $y_{j}\left(
s_{t}\right) $ is calculated. It is named \textquotedblleft neural
network\textquotedblright \ because the layers contain many units of simple
numerical operations (e.g., convolution and zero-truncation), each of which
transmits inputs and outputs in a network-like architecture, with the
analogy of computational nodes as biological neurons that transmit electric
signals.

The approximate number of parameters in AlphaGo's policy network is%
\begin{equation*}
4.6\text{ million (weights)}=\left( 192\text{ kernels}\right) ^{2}\times
\left( 5^{2}+3^{2}\times 11+1^{2}\right) ,
\end{equation*}%
where each of the \textquotedblleft kernels\textquotedblright \ is a $%
3\times 3$\ (or $5\times 5$) matrix that is designed to indicate the
presence or absence of a particular local pattern, in each $3\times 3$\ (or $%
5\times 5$)\ part of the board. Note the number of parameters of AlphaGo's
policy function\ $\left \vert \mathcal{\psi }\right \vert $ is smaller than
the 50 million parameters in Bonanza, despite the fact that Go has a larger
state space than shogi.

The supervised learning (i.e., estimation) of $\psi $ uses a standard
numerical optimization algorithm to maximize the likelihood function that
aggregates the optimal choice probabilities implied by the model, the data,
and the parameter values. That is, AlphaGo's policy function is estimated by
the classical maximum likelihood method. The team did not add any
\textquotedblleft regularization\textquotedblright \ term in the objective
function, which is a common practice in machine learning to improve the
out-of-sample prediction accuracy at the expense of biased estimates.
Nevertheless, the estimated policy function, $\sigma \left( s_{t};\hat{\psi}%
\right) $, could predict 55.7\% of the human players' moves outside the
sample, and its top-five move predictions contained the actual human choices
almost 90\% of the time.\footnote{%
By contrast, a simple parametric (logit without a DNN inside) version of the
empirical policy function (for the MCTS purposes) achieved only 27\%
accuracy, which is still remarkable but less impressive than the DNN
version's performance.}

\subsubsection*{Component 2: Reinforcement Learning (RL) of the Policy
Network}

The ultimate goal of the AlphaGo team was the creation of a strong AI, not
the prediction of human play \textit{per se} (or the unbiased estimation of $%
\psi $). The second ingredient of AlphaGo is the process of reinforcement
learning to make a stronger policy function than the estimated one from the
previous step, $\sigma \left( s_{t};\hat{\psi}\right) $.

Reinforcement learning is a generic term to describe a numerical search for
\textquotedblleft better\textquotedblright \ actions based on some
performance criteria, or \textquotedblleft reward,\textquotedblright \ such
as the average score of the game. The specific task in the current case is
to find some $\tilde{\psi}\neq \hat{\psi}$ such that the winning probability
is higher under strategy $\sigma \left( s_{t};\tilde{\psi}\right) $ than $%
\sigma \left( s_{t};\hat{\psi}\right) $:%
\begin{equation}
\Pr_{win}\left( \sigma \left( s_{t};\tilde{\psi}\right) ,\sigma \left( s_{t};%
\hat{\psi}\right) \right) >\Pr_{win}\left( \sigma \left( s_{t};\hat{\psi}%
\right) ,\sigma \left( s_{t};\hat{\psi}\right) \right) ,
\label{eq - V(tilde) > V(hat)}
\end{equation}%
where $\Pr_{win}\left( \sigma _{i},\sigma _{-i}\right) $ is the probability
that player $i$ wins with strategy $\sigma _{i}$ against the opponent who
uses $\sigma _{-i}$, in terms of the average performance across many
simulated plays of the game.

Because the outcome of the game depends on both $\sigma _{i}$ and $\sigma
_{-i}$, condition (\ref{eq - V(tilde) > V(hat)}) does not guarantee the
superiority of $\sigma \left( s_{t};\tilde{\psi}\right) $ in general (i.e.,
when playing against any strategies other than $\sigma \left( s_{t};\hat{\psi%
}\right) $). The only way to completely address this issue is to solve the
game exactly for the optimal strategy $\sigma ^{\ast }\left( s_{t}\right) $,
but such a solution is computationally impossible. Accordingly, the
development team tries to find \textquotedblleft
satisficing\textquotedblright \ $\tilde{\psi}$, by making each candidate
policy play against many different policies that are randomly sampled from
the previous rounds of iteration (i.e., various perturbed versions of $\hat{%
\psi}$ in the numerical search process), and by simulating plays from a wide
variety of $s_{t}$ that are also randomly sampled from those in the data (as
well as those from perturbed versions of such historical games).

\subsubsection*{Component 3: SL/RL of Value Network}

The third ingredient of AlphaGo is the evaluation function, $V\left(
s_{t};\theta \right) $, to assess the probability of winning from any $s_{t}$%
. Constructing such an object had been deemed impossible in the computer Go
community, but the team managed to construct the value function from the
policy function through simulations. Specifically, they proceeded as follows:

\begin{itemize}
\item Simulate many game plays between the RL policy function, $\sigma
\left( s_{t};\tilde{\psi}\right) $, and itself.

\item Pick many (30 million) different states from separate games in the
simulation and record their winners, which generates a synthetic dataset of
30 million $\left( win/loss,s_{t}\right) $ pairs.

\item Use this dataset to find the value function that predicts $\Pr_{win}$
from any $s_{t}$.
\end{itemize}

In other words, strategy $\sigma \left( s_{t};\tilde{\psi}\right) $ implies
certain outcomes in the game against itself, and these outcomes become
explicit through simulations:%
\begin{equation}
\Pr_{win}\left( \sigma \left( s_{t};\tilde{\psi}\right) ,\sigma \left( s_{t};%
\tilde{\psi}\right) \right) .
\end{equation}%
Once the outcomes become explicit, the only remaining task is to fit some
flexible functional form to predict $\Pr_{win}$ as a function of $s_{t}$,
which is a plain-vanilla regression (supervised learning) task.

The developers prepared another DNN (CNN) with a design that is similar to
the policy network: 49 channels, 15 layers, and 192 kernels. The only
differences are an additional state variable that reflects the identity of
the player (to attribute the win/loss outcome to the correct player) and an
additional computational step at the end of the hierarchical architecture
that takes all arrays of intermediate results as inputs and returns a scalar
($\Pr_{win}$) as an output.

Let us denote the estimated value network by%
\begin{equation}
V\left( s_{t};\hat{\theta},\sigma _{i}=\sigma _{-i}=\sigma \left( s_{t};%
\tilde{\psi}\right) \right) ,  \label{eq - RL value}
\end{equation}%
where the expression $\sigma _{i}=\sigma _{-i}=\sigma \left( s_{t};\tilde{%
\psi}\right) $ clarifies the dependence of $V\left( \cdot \right) $ on the
use of specific strategy by both the focal player and the opponent in the
simulation step to calculate $\Pr_{win}$. These notations are cumbersome but
help us keep track of the exact nature of the estimated value network when
we proceed to their structural interpretation.

\subsubsection*{Component 4: Combining Policy and Value with MCTS}

The fourth component of AlphaGo is MCTS, the stochastic solution method for
a large game tree. When AlphaGo plays actual games, it combines the RL
policy $\sigma \left( s_{t};\tilde{\psi}\right) $ and the RL value (equation %
\ref{eq - RL value}) within an MCTS algorithm.

Each of these components can be used individually, or in any combinations
(Silver et al. [2016], Figure 4). The policy function directly proposes the
optimal move from any given state. The value function indirectly suggests
the optimal move by comparing the winning probabilities across the next
states that result from candidate moves. An MCTS can perform a similar
state-evaluation task by simulating the outcomes of the candidate moves.
They represent more or less the same concept: approximate solutions to an
intractable problem that is guaranteed to have a unique exact solution.
Nevertheless, positive \textquotedblleft ensemble effects\textquotedblright \
from combining multiple methods are frequently reported, presumably because
different types of numerical approximation errors may cancel out each other.%
\footnote{%
This ensemble part involves many implementation details of purely
computational tasks, and hence is beyond the scope of this paper.}

\subsection{Structural Interpretation: AlphaGo Is Two-step Estimation}

\subsubsection*{Component 1: SL Policy Network Is First-Stage CCP Estimates}

Deep Blue and Bonanza embody parametric value functions, whereas AlphaGo's
first key component (SL policy network) is a nonparametric policy function,
which is equivalent to the estimation of CCPs in Hotz and Miller's (1993)
two-step method.

In case the reader is unfamiliar with the literature on dynamic structural
models, I provide a brief summary. The NFXP method (see section 4.2)
requires the solution of a DP problem, which becomes computationally
expensive as the size of the state space increases. Moreover, this solution
step has to be repeated for each candidate vector of parameter values $%
\theta $.

Hotz and Miller (1993) proposed an estimation approach to circumvent this
problem. To the extent that the actual choices in the data reflect the
optimal choice probabilities that are conditional on the observed state in
the data, we can estimate the policy function directly from the data. This
procedure is the estimation of CCPs in their first stage. The benefit of
this approach is that the procedure does not require solving a fully dynamic
model. The cost of this approach is that the requirement for data becomes
more demanding.

One should avoid imposing parametric assumptions on the first-stage policy
function, because a typical goal of empirical analysis is to find the
parameters of the value function (and its underlying structural components,
e.g., preference and technology), $\theta $, that are implied by observed
actions in data and not by their parametric surrogates. A priori
restrictions on the policy function could potentially contradict the
solution of the underlying DP. Thus, being nonparametric and preserving
flexible functional forms are crucial for an adequate implementation of the
Hotz-Miller method. In this sense, the CCP method is demanding of data. It
trades the \textit{computational} curse of dimensionality for the \textit{%
data} curse of dimensionality.

Given this econometric context, the use of DNN seems a sensible choice of
functional form. In one of the foundational works for deep learning,
econometrician Halbert White and his coauthors proved such a multi-layer
model with many nodes can approximate any arbitrary functions (Hornik,
Stinchcombe, and White [1989]), as long as the network is sufficiently large
and deep (i.e., has a sufficient degree of flexibility) to capture
complicated data patterns.\footnote{%
See also Chen and White (1999). For an overview on deep learning, see
Goodfellow, Bengio, and Courville (2016), for example.} The sole purpose and
requirement of Hotz and Miller's first stage is to capture the actual choice
patterns in the data as flexibly as possible. Silver et al. (2016) report SL
policy network's out-of-sample move-prediction accuracy is 55\% (and close
to 90\% with top-five predictions in Maddison et al. [2015]), whereas that
of a simple parametric (logit) version is 27\%. This level of fit is a
remarkable achievement, because the sample size is small (practically zero)
relative to the size of the state space.

\subsubsection*{Component 2: RL Policy Network Is Like a \textquotedblleft
Counterfactual\textquotedblright \ with Long-Lived Players}

The making of the RL policy network does not involve raw data. Rather, it is
a pure numerical search for a better approximation of the truly optimal
solution of the game (which is known to exist and be unique). Although this
paper focuses on the original version of AlphaGo, discussing its pure-RL
version (AlphaGo Zero) might be useful for a comparison:

\begin{itemize}
\item In the case of (original) AlphaGo, RL\ starts from the top human
players' strategy $\sigma \left( s_{t};\hat{\psi}\right) $ as an initial
value, and iteratively searches for a \textquotedblleft
stronger\textquotedblright \ strategy $\sigma \left( s_{t};\tilde{\psi}%
\right) $.

\item In the case of AlaphGo Zero, RL starts from \textit{tabula rasa}
(i.e., nothing but purely random play).\footnote{%
See Silver et al. (2017).}
\end{itemize}

Regardless of the choice of the initial value, RL in this context is a
policy-function iteration (or best-response iteration) approach that has
been used in many economic applications, such as Pakes and McGuire's (1994,
2001) implementation of dynamic oligopoly games.

In the case of (original) AlphaGo, the resulting strategy $\sigma \left(
s_{t};\tilde{\psi}\right) $ could be interpreted as an outcome of some
\textquotedblleft counterfactual\textquotedblright \ experiment in which the
top human players (as embodied and immortalized in $\hat{\psi}$) lived long
careers and accumulated additional experience. By the same token, the RL for
AlaphGo Zero has a flavor of simulating the learning trajectory of a
first-time player without any teacher or textbook, although the exact form
of human learning from the actual games and training would be quite
different.

\subsubsection*{Component 3: SL/RL Value Network Is Second-Stage CCS
Estimates}

According to Silver et al. (2016), AlphaGo's SL/RL value network is the
first successful evaluation function for Go, which is a remarkable
achievement. The procedure to obtain this value function is a
straightforward application of Hotz, Miller, Sanders, and Smith's (1994,
henceforth, HMSS) CCS estimator, combined with another DNN to approximate
the complicated relationship between $\Pr_{win}$ and $s_{t}$ in the
high-dimensional state space.

The literature context is the following. Hotz and Miller (1993) proved the
existence of a one-to-one mapping between the policy function and the value
function, so that the former can be inverted to estimate the latter. This
procedure is implemented by means of matrix inversion in the second stage of
their original method. However, this procedure requires the inversion of a
large matrix, the size of which increases with $\left \vert \mathcal{S}%
\right \vert $, and poses computational problems for the actual
implementation.

HMSS (1994) proposed an alternative approach to Hotz and Miller's second
step. They suggest running many forward simulations based on the first-stage
CCPs. With sufficiently many simulations, the implied value function and its
underlying structural parameters can be estimated. This principle underlies
AlphaGo's success in constructing a useful evaluation function for Go.

Although AlphaGo is developed for the game of Go and therefore a dynamic
game, its development goal is a \textquotedblleft strong\textquotedblright \
program to beat human champions (and not about studying the strategic
interactions among multiple human players in the data). Consequently,
AlphaGo's connection to the empirical dynamic-game methods seems limited.
For example, Bajari, Benkard, and Levin (2007, henceforth BBL) extended HMSS
(1994) to dynamic games and proposed a moment-inequality-based estimation
approach. The development process of game AIs (including AlphaGo) typically
abstracts from strategic interactions.\footnote{%
Section 6 discusses this and other related issues.}

\subsubsection*{Component 4: MCTS and Ensemble}

The actual play of AlphaGo is generated by a complex combination of the
estimated/reinforced policy function, the value function, and MCTS.

The MCTS part involves randomly playing out many games. This
\textquotedblleft random\textquotedblright \ play is generated from a simple
version of the empirical policy function that resembles a standard logit
form (i.e., one without the multi-layer architecture of DNN to compute $%
y_{j}\left( s_{t}\right) $ as in equation \ref{eq - logit}). Hence, AlphaGo
in the actual game is a hybrid of the following:

\begin{itemize}
\item the reinforced version of top human players' strategy (as represented
in a deep, convolutional logit functional form),

\item their implicit value function (with a similar DNN specification), and

\item real-time forward simulation based on the estimated \textquotedblleft
quick-and-dirty\textquotedblright \ policy function (in a simple logit form).
\end{itemize}

\subsubsection*{AlphaGo Zero: All-in-One Package}

The new version of the program published in 2017, AlphaGo Zero (Silver et
al. [2017]), does not use any human data (or handcrafted features to
represent the board state). Because it has no empirical (human data-related)
component, this new AI is not an obvious subject in terms of econometrics.%
\footnote{%
See paragraphs on RL policy function in the earlier part of this section.}

Nevertheless, one aspect of its architectural design seems intriguing: a
single neural network to perform the functions of both the policy network
and value network in the original version. This single network is larger
than the previous networks and is combined with MCTS for purely
simulation-based search for the optimal strategy to play Go.

This design change is reasonable from the perspective of economic modeling,
because the construction of all three of the policy network, value network,
and MCTS algorithm in the original AlphaGo was conceptually redundant.
Policy and value are dual objects; one implies the other. Likewise, MCTS is
a search for the optimal strategy on its own. The \textquotedblleft
ensemble\textquotedblright \ effect from combining the three components might
have conferred an additional performance gain, but it is ultimately a
manifestation of approximation errors within individual components.

Finally, much of the performance gains in AlphaGo Zero seem to stem from the
significantly larger size of the neural network architecture (i.e., a more
flexible functional form to parameterize the state space). Hence, its
superior performance against the original version does not necessarily speak
to the costs and benefits of using human data by themselves.

\section{Implicit Assumptions}

Sections 3, 4, and 5 explained how the concepts and algorithms behind the
three game AIs correspond to more familiar ideas and methods in the
economics of dynamic structural models.

\begin{itemize}
\item Deep Blue is a calibration of a linear value function.

\item Bonanza's machine-learning method\ is equivalent to Rust's (1987) NFXP
algorithm.

\item AlphaGo's SL policy network\ is a DNN\ implementation of Hotz and
Miller's (1993) first-stage nonparametric CCP estimator.

\item AlphaGo's SL/RL value network\ is a DNN\ implementation of HMSS's
(1994) second-stage CCS estimator.
\end{itemize}

This section clarifies some of the implicit assumptions underlying (the
algorithms to develop) these AIs, and discusses their implications.

Deep Blue does not use data analysis (at least for its main component), but
both Bonanza and AlphaGo use logit-style discrete-choice models, and
therefore implicitly assume the presence of an error term associated with
each of the available actions $a_{t}\in \mathcal{A}\left( s_{t}\right) $: $%
\varepsilon \left( a_{t}\right) \sim $ type-1 extreme value.\footnote{%
Because the goal of these games is to win, $\varepsilon \left( a_{t}\right) $
does not contribute to the eventual payoff $u_{i}\left( s_{t}\right) $. The
AIs' formulation treats $\varepsilon \left( a_{t}\right) $ as a purely
transient, random component of payoffs that players care about only at the
time of choosing concurrent $a_{t}$.} The inclusion of this continuous
random variable to each discrete alternative eliminates the possibility of
ties between the payoffs of multiple actions $j$ and $j^{\prime }$, thereby
making the mapping between the value function and the policy function unique.

In the plain-vanilla application of discrete-choice models, including
Bonanza and AlphaGo, this error term is assumed i.i.d. across actions,
players, time, and games. In some empirical contexts, however, one might
want to consider relaxing this distributional assumption.

\textbf{(1) Consideration Sets and Selective Search }For example, a subset
of legal moves $\mathcal{C\subset A}\left( s_{t}\right) $ could belong to 
\textit{joseki}, or commonly known move patterns, whereas other moves are
not even considered by human experts (i.e., outside their \textquotedblleft
consideration sets\textquotedblright ) unless new research shows their
effectiveness. Similarly, experts are forward-looking but focus on only a
few moves per decision node, conducting a highly selective game-tree search.
Capturing these aspects of human play would require the econometrician to
distinguish between choice sets and consideration sets.

\textbf{(2) Cross-sectional Heterogeneity }Large $\left \vert \mathcal{A}%
\right \vert $, $\left \vert \mathcal{\theta }\right \vert $, and $%
\left
\vert \mathcal{\psi }\right \vert $ necessitate the data-analysis
part of the AIs to pool data from all games regardless of the identity of
players or occasions. However, players are heterogeneous in their styles and
strengths. Such differences would become a source of systematic
heterogeneity in $\varepsilon \left( a_{t}\right) $ across players and
contradict the i.i.d. assumption.

\textbf{(3) Inter-temporal Heterogeneity }Likewise, the state of knowledge
about desirable moves, or \textit{joseki}, evolves over time as a result of
new games, experimentation, and research (including the emergence of game
AIs and their play styles). The evolving nature of knowledge and strategies
becomes a source of systematic heterogeneity in $\varepsilon \left(
a_{t}\right) $ across time.

\textbf{(4) Strategic Interactions }Each game in the data embodies two
specific players' attempts to out-maneuver each other. When experts prepare
for games, they study rival players' past and recent strategies to form
beliefs about their choice probabilities and exploit any weaknesses. Such
interactions continue during the actual games, as players keep updating
their beliefs and adjust strategies accordingly. By contrast, both the
development process and the actual play of game AIs are based on the
assumption that $\sigma _{-i}=\sigma _{i}$ (and more or less equivalently, $%
V_{-i}=V_{i}$), abstracting from strategic interactions.

\textbf{(5) Time and Physical Constraints }The exposition so far has mostly
abstracted from the notion of time constraints, but official games impose
limits on the amount of time each player can spend on thinking. This time
constraint makes the game nonstationary in terms of clock time, and adds
another dimension to the player's optimization problem in terms of the
intertemporal allocation of thinking time. The data analysis for the game
AIs abstracts from these fundamental aspects of human play, although
computers face the same constraint in actual games. For AIs, time
constraints manifest themselves either as the length $L$ of a truncated game
tree (in the case of Deep Blue and Bonanza) or as the number of play-out
simulations for MCTS (in the case of AlphaGo).

Similar constraints exist in terms of physical (or mental) capacity in terms
of computation speed, information storage, and precision of these
operations. Human players are more constrained than computers and more prone
to obvious mistakes, especially under severe time constraints. In fact, an
important aspect of strategic interactions between human experts is about
encouraging the opponent to make mistakes. Mistakes would make the implicit $%
\varepsilon \left( a_{t}\right) $ irregular, and explicitly incorporating
time and physical constraints would lead to different modeling approaches.

\section{Opportunities for Future Research}

\subsubsection*{Relaxing the Implicit Assumptions to Capture Human Behavior}

Relaxing these implicit assumptions (in the previous section) would make the
underlying models of human behavior for game AIs more realistic (i.e., more
human-like). Having achieved performance milestones in terms of pure
strength (i.e., approximating the optimal strategy better than top human
players), approximating human behavior could be one of the renewed research
goals for game AI development. Adding ad hoc features to emulate humans is
one way, but developing and estimating a more realistic model could be
another, perhaps more fundamental approach.

The econometrics of dynamic structural models have advanced considerably
since the time of Rust (1987), Hotz and Miller (1993), and HMSS (1994) to
address the issues in section 6. Incorporating various kinds of
heterogeneity as well as analyzing strategic interactions has been central
to this progress.\footnote{%
Bajari, Benkard, and Levin (2007), Aguirregabiria and Mira (2007), Pakes,
Ostrovsky, and Berry (2007), and Pesendorfer and Schmidt-Dengler (2008)
proposed methods for analyzing dynamic games along the lines of the two-step
estimation method, whereas recent empirical applications, such as Igami
(2017, 2018), Zheng (2016), Yang (2017), and Igami and Uetake (2017), build
on the full-solution method.
\par
Kasahara and Shimotsu (2009) propose a method (based on rank conditions of
the state transition dynamics) to identify the lower bound of the number of
unobserved types that is required to rationalize data patterns. Arcidiacono
and Miller (2011) use an expectation-maximization algorithm to estimate CCPs
in the presence of such unobserved types. Berry and Compiani (2017) advance
an instrumental-variables approach to address unobserved heterogeneity in
dynamic models.} Such new methods can be applied to the task of making the
AIs' underlying models more realistic.

\subsubsection*{Structural Econometrics for \textquotedblleft Explainable
AI\textquotedblright}

Relaxing the implicit econometric assumptions would make the models not only
more realistic, but also more interpretable. One of the benefits of
developing and estimating a structural model is that the results are
economically interpretable, above and beyond the basic notions of causation
and correlation in simpler settings (e.g., determining a statistical
relationship between some variables $X$ and $Y$). The words
\textquotedblleft interpretable\textquotedblright \ and \textquotedblleft
explainable\textquotedblright \ could mean different things in different
fields, but the concept of \textquotedblleft structural
interpretability\textquotedblright \ seems useful as a guide for a more
formal definition.

Note this proposal about structural interpretation should not be confused
with the challenge concerning \textquotedblleft explaining
DNNs.\textquotedblright \ DNNs are a flexible functional form, or a class of
basis functions for nonparametric estimation, and therefore do not have
economic interpretation by themselves. By contrast, the object for which
these functional forms are specified could have a structural interpretation
(e.g., AlphaGo's SL policy network is a CCP estimate of the average
professional player's strategy under the maintained assumptions of
homogeneity etc.).

\subsubsection*{DNN for Nonparametric CCP Estimation}

The use of DNN specifications for the first-stage nonparametric estimation
of CCPs seems a good idea. This class of model specification has long been
known to be capable of approximating arbitrary functions, but AlphaGo offers
a proof of concept in the dynamic-game context, which is sufficiently
complicated and potentially relevant for economic applications.

\bigskip

Clarifying the mapping between the two fields is only a first step toward
cross-fertilization, but the opportunities for future research seem to
suggest themselves.

\section*{Appendix: \ Functional-Form Specification of AlphaGo}

AlphaGo's policy function uses the following functional form, and its value
function has a similar architecture. It consists of 48 input
\textquotedblleft channels\textquotedblright \ (variables), 13
\textquotedblleft layers\textquotedblright \ (stages within a hierarchical
architecture), and 192 \textquotedblleft kernels\textquotedblright \ (filters
to find local patterns). A complete review of deep neural networks in
general (or AlphaGo's model specification in particular) is beyond the scope
of this paper, but these objects interact as follows. Each of the 48
channels represents a binary indicator variable that characterizes $s_{t}$:%
\begin{equation}
x_{kt}=\left \{ 
\begin{array}{ll}
1 & \text{if feature }k\text{\ is present in }s_{t},\text{ and} \\ 
0 & \text{otherwise.}%
\end{array}%
\right. 
\end{equation}%
\textquotedblleft Features\textquotedblright \ include the positions of black
stones, white stones, and blanks (see Extended Data Table 2 of Silver et al.
[2016] for the full list).

These $x_{kt}$'s are not combined linearly (as in $V_{DB}$ or $V_{BO}$) but
are processed by many kernels across multiple hierarchical layers. In the
first layer, each of the 192 kernels is a $5\times 5$ grid with 25
parameters that responds to a particular pattern within 25 adjacent
locations.\footnote{%
The \textquotedblleft patterns\textquotedblright \ that these kernels are
designed to pick up should be distinguished from the initial 48
\textquotedblleft features\textquotedblright \ (variables) representing the
board state in the original input data.} As the name \textquotedblleft
kernel\textquotedblright \ suggests, this $5\times 5$ matrix is applied to
perform convolution operations at every one of the $225$ $\left( =15\times
15\right) $ locations within the $19\times 19$ board:%
\begin{equation}
z_{r,c}=\sum_{l=1}^{192}\sum_{p=1}^{5}\sum_{q=1}^{5}w_{l,p,q}\times
x_{l,r+p,c+q}+b,  \label{eq- convolution}
\end{equation}%
where $z_{r,c}$ is the result of convolution for row $r$ and column $c$, $%
w_{l,r,q}$ is the weight for kernel $l$, row $r$, and column $c$, $\left(
p,q\right) $ denote the row and column of the kernel, $x_{l,r+p,c+q}$\ is an
input, and $b$ is the intercept term (\textquotedblleft
bias\textquotedblright ). The weights and intercepts constitute the
parameters of the model (collectively denoted by $\psi $ in the main text).
DNNs of this type are called convolutional neural networks (CNNs) in the
machine-learning literature, and are primarily used for image-recognition
tasks.\ The results of convolution are subsequently transformed by a
function,%
\begin{equation}
y_{r,c}=\max \left \{ 0,z_{r,c}\right \} ,  \label{eq - ReLU}
\end{equation}%
where $y_{r,c}$ is the transformed output (to be passed on to the next layer
as input). This function is called rectified linear unit (or
\textquotedblleft ReLU\textquotedblright ). The resulting $15\times 15$
output is smaller than the $19\times 19$ board; the margins are filled by
zeros to preserve the $19\times 19$ dimensionality (\textquotedblleft zero
padding\textquotedblright ).

In the second layer, another set of 192 kernels is used to perform
convolution on the outputs from the first layer. The results go through the
ReLU transformation again, and proceed to the third layer. The size of the
kernels in layers 2 through 12 is $3\times 3$, instead of $5\times 5$ in
layer 1. In layer 13, the size of the layer is $1\times 1$, because the goal
of the policy network is to put a number on each of the $19\times 19$ board
positions without \textquotedblleft zero padding\textquotedblright \ the
margins. Each of the $19\times 19$ outputs in this last layer goes through a
logit-style monotonic (\textquotedblleft softmax\textquotedblright )
transformation into the $\left[ 0,1\right] $ interval,%
\begin{equation}
CCP_{r,c}=\frac{\exp \left( y_{r,c}\right) }{\sum_{r^{\prime
}}\sum_{c^{\prime }}\exp \left( y_{r^{\prime },c^{\prime }}\right) },
\label{eq - softmax}
\end{equation}%
so that the final output can be interpreted as the players' conditional
choice probabilities of choosing action $j$ (or board location $\left(
r,c\right) $).


\begin{thebibliography}{99}
\bibitem{} Arcidiacono, P., and R. A. Miller. 2011. \textquotedblleft
Conditional Choice Probability Estimation of Dynamic Discrete Choice Models
With Unobserved Heterogeneity.\textquotedblright \  \emph{Econometrica}, 79:
1823--1867.

\bibitem{} Athey, Susan. 2017. \textquotedblleft Beyond prediction: Using
big data for policy problems.\textquotedblright \  \emph{Science}, 355:
483--485.

\bibitem{} Bajari, P., C. L. Benkard, and J. Levin. 2007 \textquotedblleft
Estimating Dynamic Models of Imperfect Competition.\textquotedblright \ 
\emph{Econometrica}, 75: 1331--1370.

\bibitem{} Belloni, A., V. Chernozhukov, and C. Hansen. 2014.
\textquotedblleft High-Dimensional Methods and Inference on Structural and
Treatment Effects.\textquotedblright \  \emph{Journal of Economic Perspectives%
}, 28: 29--50.

\bibitem{} Berry, S. T., and G. Compiani. 2017. \textquotedblleft An
Instrumental Variable Approach to Dynamic Models.\textquotedblright \
Manuscript, Yale University.

\bibitem{} Campbell, M., A. Hoane, and F. Hsu. 2002. \textquotedblleft Deep
Blue.\textquotedblright \  \emph{Artificial Intelligence}, 134: 57--83.

\bibitem{} Chen, Xiaohong, and Halbert White. 1999. \textquotedblleft
Improved Rates and Asymptotic Normality for Nonparametric Neural Network
Estimators.\textquotedblright \  \emph{IEEE Transactions on Information Theory%
}, 45 (2): 682--691.

\bibitem{} Goodfellow, I., Y. Bengio, and A. Courville. 2016. \emph{Deep
Learning}. Cambridge, MA: The MIT Press.

\bibitem{} Habu, Yoshiharu, and NHK. 2017. \emph{Jinkou chinou no kakushin}.
Tokyo: NHK shuppan.

\bibitem{} Hoki, K. 2012. \textquotedblleft Kazuno bouryoku de ningen ni
chousen! Bonanza no tanjou,\textquotedblright \ in Computer Shogi
Association, ed., \emph{Ningen ni katsu computer shogi no tsukuri kata}.
Tokyo: Gijutsu hyouron sha.

\bibitem{} Hoki, K., and A. Watanabe. 2007. \emph{Bonanza Vs Shoubunou:
Saikyou shogi sohuto wa ningen wo koeruka}. Tokyo: Kadokawa (in Japanese).

\bibitem{} Hornik, K., M. Stinchcombe, and H. White. 1989. \textquotedblleft
Multilayer Feedforward Networks are Universal
Approximators,\textquotedblright \  \emph{Neural Networks}, 2: 359--366.

\bibitem{} Hotz, V. J., and R. A. Miller. 1993. \textquotedblleft
Conditional Choice Probabilities and the Estimation of Dynamic
Models.\textquotedblright \  \emph{Review of Economic Studies}, 60: 497--529.

\bibitem{} Hotz, V. J., R. A. Miller, S. Sanders, and J. Smith. 1994.
\textquotedblleft A Simulation Estimator for Dynamic Models of Discrete
Choice.\textquotedblright \  \emph{Review of Economic Studies}, 61: 265--289.

\bibitem{} Igami, M.. 2017. \textquotedblleft Estimating the Innovator's
Dilemma: Structural Analysis of Creative Destruction in the Hard Disk Drive
Industry, 1981--1998,\textquotedblright \  \emph{Journal of Political Economy}%
, 125: 798--847.

\bibitem{} Igami, M.. 2018 \textquotedblleft Industry Dynamics of
Offshoring: The Case of Hard Disk Drives.\textquotedblright \  \emph{American
Economic Journal: Microeconomics}, forthcoming.

\bibitem{} Igami, M., and K. Uetake. 2017. \textquotedblleft Mergers,
Innovation, and Entry-Exit Dynamics: Consolidation of the Hard Disk Drive
Industry, 1996--2016.\textquotedblright \ Manuscript, Yale University.

\bibitem{} Kaneko, Tomoyuki. 2012. \textquotedblleft GPS Shogi no
tanjou,\textquotedblright \ in Computer Shogi Association, ed., \emph{Ningen
ni katsu computer shogi no tsukuri kata}. Tokyo: Gijutsu hyouron sha.

\bibitem{} Kasahara, H., and K. Shimotsu. 2009. \textquotedblleft
Nonparametric Identification of Finite Mixture Models of Dynamic Discrete
Choices.\textquotedblright \  \emph{Econometrica}, 77: 135--175.

\bibitem{} Kasparov, G.. 2007. \emph{How Life Imitates Chess: Making the
Right Moves, from the Board to the Boardroom}. London: Bloomsbury.

\bibitem{} Maddison, C. J., A. Huang, I. Sutskever, and D. Silver. 2015.
\textquotedblleft Move Evaluation in Go Using Deep Convolutional Neural
Networks.\textquotedblright \  \emph{ICLR}.

\bibitem{} Mullainathan, S., and J. Spiess. 2017. \textquotedblleft Machine
learning: an applied econometric approach.\textquotedblright \  \emph{Journal
of Economic Perspectives}, 31: 87--106.

\bibitem{} Otsuki, T.. \emph{Saikyou igo AI AlphaGo kaitai shinsho}. Tokyo:
Shoeisha (in Japanese).

\bibitem{} Pakes, A., and P. McGuire. 1994. \textquotedblleft Computing
Markov-Perfect Nash Equilibria: Numerical Implications of a Dynamic
Differentiated Product Model.\textquotedblright \  \emph{RAND Journal of
Economics}, 25 (4): 555--589.

\bibitem{} Pakes, A., and P. McGuire. 2001. \textquotedblleft Stochastic
Algorithms, Symmetric Markov Perfect Equilibrium, and the `curse' of
Dimensionality.\textquotedblright \  \emph{Econometrica}, 69 (5): 1261--1281.

\bibitem{} Pakes, A., Ostrovsky, M., and Berry, S. 2007. \textquotedblleft
Simple estimators for the parameters of discrete dynamic games (with
entry/exit examples).\textquotedblright \  \emph{RAND Journal of Economics},
38: 373--399.

\bibitem{} Pesendorfer, M., and P. Schmidt-Dengler. 2008. \textquotedblleft
Asymptotic Least Squares Estimators for Dynamic Games.\textquotedblright \ 
\emph{Review of Economic Studies}, 75: 901--928.

\bibitem{} Rust, J.. 1987. \textquotedblleft Optimal Replacement of GMC Bus
Engines: An Empirical Model of Harold Zurcher.\textquotedblright \  \emph{%
Econometrica}, 55: 999--1033.

\bibitem{} Rust, J.. 2017. \textquotedblleft Dynamic Programming,
Numerical.\textquotedblright \  \emph{Wiley StatsRef: Statistics Reference
Online}.

\bibitem{} Silver, D., A. Huang, C. J. Maddison, A. Guez, L. Sifre, G. van
den Driessche, J. Schrittwieser, I. Antonoglou, V. Panneershelvam, M.
Lanctot, S. Dieleman, D. Grewe, J. Nham, N. Kalchbrenner, I. Sutskever, T.
Lillicrap, M. Leach, K. Kavukcuoglu, T. Graepel, and D. Hassabis. 2016.
\textquotedblleft Mastering the game of Go with deep neural networks and
tree search.\textquotedblright \  \emph{Nature}, 529: 484--489.

\bibitem{} Silver, D., J. Schrittwieser, K. Simonyan, I. Antonoglou, A.
Huang, A. Guez, T. Hubert, L.~Baker, M. Lai, A.~Bolton, Y.~Chen, T.
Lillicrap, F. Hui, L. Sifre, G. van den Driessche, T. Graepel, and D.
Hassabis. 2017. \textquotedblleft Mastering the game of Go without human
knowledge.\textquotedblright \  \emph{Nature}, 550: 354--359.

\bibitem{} Tsuruoka, Yoshimasa. 2012, \textquotedblleft Gekisashi no
tanjou,\textquotedblright \ in Computer Shogi Association, ed., \emph{Ningen
ni katsu computer shogi no tsukuri kata}. Tokyo: Gijutsu hyouron sha.

\bibitem{} Varian, Hal. 2014. \textquotedblleft Big Data: New Tricks for
Econometrics.\textquotedblright \  \emph{Journal of Economic Perspectives},
28: 3--28.

\bibitem{} Watanabe, Akira. 2013. \textit{Sh\={o}bushin}. Tokyo: Bungei
shunju (in Japanese).

\bibitem{} Watanabe, Akira. 2014. \textit{Watanabe Akira no shikou: Banjou
bangai mondou}. Tokyo: Kawade shobou shinsha (in Japanese).

\bibitem{} Yamamoto, I.. 2017. \emph{Jinkou chinou wa donoyouni shite
\textquotedblleft Meijin\textquotedblright \ wo koetanoka?} Tokyo: Diamond
sha (in Japanese).

\bibitem{} Yang, Chenyu. 2017. \textquotedblleft Could Vertical Integration
Increase Innovation?\textquotedblright \ Manuscript, University of Rochester.

\bibitem{} Yoshizoe, K., and H. Yamashita. 2012. \emph{Computer Go: Theory
and Practice of Monte Carlo Method}\ (ed. by H. Matsubara). Tokyo: Kyouritsu
shuppan (in Japanese).

\bibitem{} Zheng, Fanyin. 2016. \textquotedblleft Spatial Competition and
Preemptive Entry in the Discount Retail Industry.\textquotedblright \
Manuscript, Columbia University.
\end{thebibliography}
\end{document}